\documentstyle[aps,prb,epsf]{revtex}
\advance\textheight by 0.1in
\preprint{IASSNS-HEP-99/??}
\date{\today}
\title{
Holons on a meandering stripe}
\author{
Oleg Tchernyshyov\cite{email-oleg} and 
Leonid P.~Pryadko}
\address{School of Natural Sciences, Institute for Advanced Study, 
Princeton, New Jersey 08540}

\begin{document}
\twocolumn[\hsize\textwidth\columnwidth\hsize\csname %
  @twocolumnfalse\endcsname

\maketitle
\begin{abstract}
  To study a possible effect of transverse fluctuations of a stripe in
  a two-dimensional antiferromagnet on its charge dynamics, we
  identify elementary excitations of a weakly doped domain wall in the
  Hubbard model.  Hartree-Fock numerics and analysis of fermion zero
  modes suggest that for $U\ge3t$ charged excitations are mobile
  holons, $Q=1$, $S=0$.  Each holon resides on a kink in the position of
  the domain wall.  We construct a simple model in which transverse
  stripe dynamics is induced solely by motion of the holons.  In the
  absence of spin excitations (spinons, $Q=0$, $S=1/2$), 
  stripe fluctuations {\em do not\/} suppress a tendency to form a
  global charge-density order.
\end{abstract}
\pacs{XXX}
]

Physics of charge carriers confined to an antiphase domain wall in a
two-dimensional antiferromagnet (AF) is one of the most important and
least understood ingredients of the
stripe\cite{Tranquada-98,early-theory}-based
approach\cite{Kivelson-LosAlamos} to
high-$T_c$ superconductivity.  Both the long-distance
interactions~\cite{Kivelson-LosAlamos,Low-94} and the short-distance
physics~\cite{Zaanen-JPCS,PKEBD-99} seem to be important for understanding 
even the basic phenomenology of stripes.
Given\cite{Tranquada-98,Nayak-97,White,Hellberg-99} the
existence of metallic stripes in an antiferromagnet, their physics
should not be far different from that of a more conventional
one-dimensional electron gas (1DEG).
However, as suggested by model calculations,\cite{spin-gap} a metallic stripe
with repulsive interactions placed in an AF environment 
may develop a spin gap. 
(In a conventional spin-rotation-invariant 1DEG a spin gap can only
appear in the presence of attractive interactions.\cite{Emery-1DEG})
With spin degrees of freedom gapped, coupling between adjacent
stripes could lead either to a global superconducting order, or to an
insulating charge-density wave (CDW) state.\cite{Kivelson-LosAlamos}
In a 1DEG with repulsive interactions, CDW correlations 
diverge more strongly,\cite{Emery-1DEG} and one would expect them to
dominate superconductivity.  A way around this difficulty was
suggested\cite{Kivelson-98} by Kivelson, Fradkin and Emery (KFE), who
noticed that transverse fluctuations of conducting stripes would
induce a phase mismatch between CDWs on neighboring
stripes, effectively suppressing the interstripe coupling and the
tendency to form a charge-ordered state.  In contrast, such
fluctuations {\em do not\/} suppress superconducting ordering, since
the superconducting phase is not spatially modulated.

Transverse dynamics of insulating\cite{insulating-stripes} and
metallic\cite{Zaanen9804300} stripes has been previously studied by
Zaanen {\em et al}.  
Of a particular interest to us is the example\cite{Zaanen9804300} of a
metallic stripe with a built-in coupling of charge motion and
transverse fluctuations.  The model describes an initially insulating
stripe with a $4k_F$ CDW ($1/2$ electron per unit cell).  Assuming
that doped holes reside on rows adjacent to the site-centered stripe,
Zaanen {\em et al.\/}\ found charge-$1/2$ solitons, which shift the
domain wall one lattice constant sideways (left/right), thus acquiring
a transverse flavor.

\begin{figure}
\centering
\leavevmode
\epsfxsize 0.9\columnwidth
\epsffile{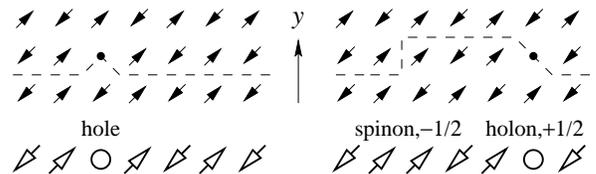}
\caption{Spin-charge separation on a bond-centered domain wall
(dashed line) in an easy-axis AF ($J_z\gg J_x,t$).  A hole decays 
into a spinon ($Q=0$, $S=1/2$) and 
a holon ($Q=1$, $S=0$) with isospins as shown (see text). 
Open symbols denote $y$-integrated spin and charge.  }
\label{fig_cartoon}
\end{figure}

In this paper, we construct a similar model to study the effect of
transverse fluctuations of a stripe on its charge dynamics.  First, we
employ Hartree-Fock (HF) numerics and an analysis of fermion zero
modes to identify elementary excitations of a lightly doped domain
wall in the Hubbard model.  Of course, such a texture is an excited,
only locally stable configuration within the HF approximation.
Nevertheless, it could be a good starting point for understanding the
nature of globally stable metallic stripes.  We find that a
weakly-doped domain wall is {\em bond\/}-centered.  At moderate and strong
coupling, $U\ge3t$, holes doped into such a wall create kinks in the
transverse position of the stripe (wiggles) and form mobile solitons
with spin $S=0$ and charge $Q=1$ (holons, Fig.~\ref{fig_cartoon}).
Unlike their 1D counterparts, holons on a domain wall have a
transverse degree of freedom, an isospin: a right (left) wiggle is
ascribed isospin $\rho=+1/2$ ($-1/2$).  In this respect, they resemble
the solitons of Zaanen {\em et al.\/}\cite{Zaanen9804300} Second, we
construct an effective 1D model of a stripe whose transverse
fluctuations are associated solely with the motion of the holons,
while spinons (kinks of a similar nature but with $Q=0$, $S=1/2$) are
frozen out.  In contrast with the KFE arguments,\cite{Kivelson-98}
transverse fluctuations of a stripe in our model {\em do not\/} break
phase coherence of CDWs on neighboring stripes.  Therefore, they {\em
do not\/} suppress the tendency to global CDW ordering, although the
effect may be restored with the inclusion of spinons.

\begin{figure}
\noindent\leavevmode\epsfxsize\columnwidth
\epsffile{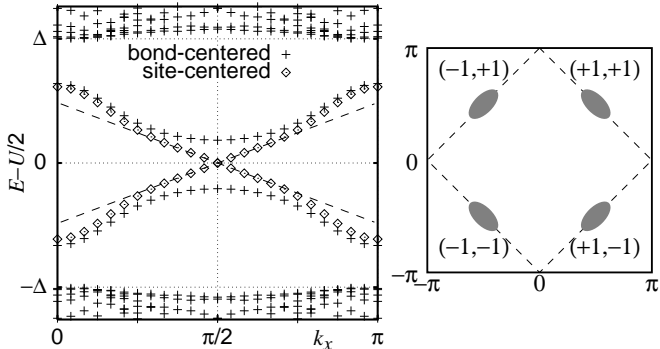}
\caption{Left: midgap one-particle spectrum $E(k_x)$ of an AF with a
  straight domain wall.  Self-consistent solution of Hartree-Fock
  equations.  $U=2.5t$, $48\!\times\!27$ sites.  Dashed line:
  low-energy approximation, Eq.~(\protect\ref{H_eff_scw}).  States
  outside the gap are not shown for the site-centered stripe.  Right: 4
  Fermi patches in the Brillouin zone, Eq.~(\protect\ref{trispinor}),
  shown with eigenvalues $(\tau^x_3,\tau^y_3)$.  Dashed line: Fermi
  surface of the noninteracting system.  }
\label{fig_dirac}
\end{figure}

{\em Numerical mean-field solutions.}  We have solved
self-consistently Hartree-Fock equations for a linearly polarized AF,
$\langle s_x \rangle = \langle s_y \rangle = 0$, $\langle s_z \rangle
\neq 0$, for small and intermediate interaction strengths
$U=2\ldots4t$ of the Hubbard model.  At half-filling, we have found
locally stable self-consistent solutions with a bond-centered domain
wall.  A site-centered wall has a slightly higher energy.  A reduction
of staggered magnetization on the wall induces a 1D band in the middle
of the Hubbard gap $\Delta$ (Fig.~\ref{fig_dirac}).  Fermions confined
to a site-centered wall have a Dirac spectrum $E = \pm v
\,|k_x-\pi/2|$ and acquire a small mass (gap) $\tilde{\Delta} <
\Delta$ in the case of a bond-centered wall.  This gap arises because
an electron moving along the wall feels a nonzero $x$-staggered
magnetization (vanishing by symmetry for a site-centered wall).  At
half-filling, the Fermi energy lies between the two midgap bands
($E_F=U/2$).

\begin{figure}
\centering
\leavevmode
\epsfxsize 0.9\columnwidth
\epsffile{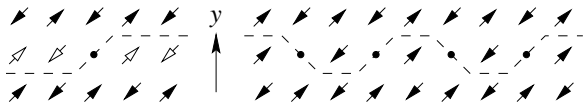}
\vskip 2mm
\caption{Left: Bond-centered stripe with a wiggle 
as a superposition of a site-centered domain wall (black arrows) and a 
1D AF chain with a kink (open arrows).  
Right: Holons with alternating isospins form a cite-centered stripe. 
}
\label{fig_site_to_bond}
\end{figure}

\begin{figure}
\centering 
\leavevmode 
\epsfxsize\columnwidth
\epsffile{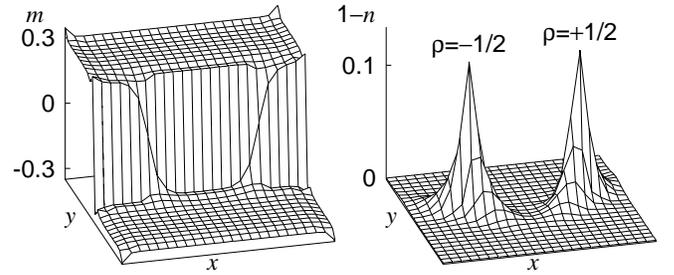} 
\caption{Staggered
magnetization $m({\bf r}) = (-1)^{x+y}\langle s_z({\bf r})\rangle$ and
hole density $1-n({\bf r})$ in a Hartree-Fock calculation at $U=3t$,
$24\!\times\!24$ sites.  A bond-centered wall with 2 wiggles and 2
doped holes.
}
\label{fig_holon}
\end{figure}

A bond-centered domain wall {\em with a wiggle\/} is similar to an AF
chain with a kink in $x$-staggered magnetization
(Fig.~\ref{fig_site_to_bond}).  As in the model of
polyacetylene,\cite{SSH} a pair of degenerate localized states (zero
modes) appears in the middle of the smaller gap.  A wall with a kink
doped with a single hole has a $S=0$, $Q=1$ soliton at the wiggle
(Fig.~\ref{fig_holon}).  In view of translational invariance, the
charged defect can move along the wall.  We thus identify it with a
holon of the large-$J_z$ limit (Fig.~\ref{fig_cartoon}).  At
$U=2\ldots3t$ (rather weak coupling), holons are cigar-shaped like
spin bags of Schrieffer {\em et al.},\cite{Schrieffer-Wen-Zhang} a
consequence of the Fermi surface nesting.  Orientation of a holon
along one or the other lattice diagonal is correlated with the
direction of the wiggle (the transverse flavor).
We have verified that holons are the lowest-energy states of doped
charges on an empty wall for $U\ge3t$ (at the HF level).  At a weaker
coupling, they bind ito bipolarons with $Q=2$, $S=0$, and zero isospin.

{\em Continuum Hartree-Fock approximation}. HF equations for a
linearly polarized Hubbard AF read
\begin{equation}
-t\sum_{\langle{\bf r\, r'}\rangle} \psi_{-s}({\bf r'}) 
+ U [\delta n({\bf r}) - s\, m({\bf r})] \psi_s({\bf r}) = 
E\psi_s({\bf r}),
\label{staggered_HF}
\end{equation}
where $m({\bf r})
= (-1)^{x+y}\langle s_z({\bf r})\rangle$ 
is the staggered magnetization and 
$s$ is the {\em staggered\/} spin index.  In what follows, 
density fluctuations $\delta n({\bf r})$ will be neglected \cite{Neglect_dn}
to restore charge-conjugation symmetry. 
Eq.~(\ref{staggered_HF}) can be cast in a matrix form 
using Pauli matrices $\{\sigma_i\}$.  
Staggered spin $s$ is an eigenvalue of $\sigma_3$, 
the hopping term is proportional to $\sigma_1$.
Most importantly, zero modes of Eq.~(\ref{staggered_HF}) are invariant under 
charge conjugation $\psi({\bf r}) \to \sigma_2\psi^*({\bf r})$.  
If such modes are present, the system contains 
solitons with $Q=0$, $S=1/2$ (spinons) or $Q=1, S=0$ 
(holons).\cite{0-holon} 

We have determined (see, e.g., Fig.~\ref{fig_dirac}, left) 
that low-energy midgap states induced by an empty domain wall  
live at one of the four ``Fermi patches'' with lattice momenta
$|k_x| \approx |k_y| \approx \pi/2$ (Fig.~\ref{fig_dirac}, right).  
Introducing smoothly varying amplitudes of a fermion 
wavefunction,
\begin{equation}
\psi_s({\bf r}) \approx 
{\textstyle \sum_{\alpha=\pm1} \sum_{\mu=\pm1}}\ 
\psi_{\alpha\mu s}({\bf r})e^{i\pi (\alpha x + \mu y)/2},
\label{trispinor}
\end{equation}
adds two more indices, $\alpha = {\rm sgn}\,k_x$ and 
$\mu = {\rm sgn}\,k_y$.  Accordingly, 
we preserve only those Fourier components of magnetization 
which connect the Fermi patches:
$$\langle s_z({\bf r})\rangle \approx 
{\textstyle \sum_{\alpha=0}^{1} \sum_{\mu=0}^{1}}\ 
m_{\alpha\mu}({\bf r})e^{i\pi (\alpha x + \mu y)}.$$
With the new indices come two more mutually
commuting sets of Pauli matrices, $\{\tau^x_i\}$ and $\{\tau^y_i\}$, 
$i=1,2,3$.  In terms of these,
${\bf k} \approx \frac{\pi}{2}(\tau^x_3, \tau^y_3)$, 
$(-1)^x = \tau^x_1$, $(-1)^y = \tau^y_1$.  
The HF Hamiltonian becomes 
\begin{eqnarray}
H_{\rm HF} = 
-2ita\,\sigma_1\tau^x_3\partial_x  -2ita\,\sigma_1\tau^y_3\partial_y
- U\sigma_3 m({\bf r}), 
\label{HF_continuum_2}
\\
m \equiv m_{11} + m_{01}\tau^x_1 
+ m_{10}\tau^y_1 + m_{00}\tau^x_1\tau^y_1. 
\nonumber
\end{eqnarray}
Only $m_{11}({\bf r})$, staggered magnetization proper, survives 
in the bulk inducing the Hubbard gap 
$\Delta = U|m_{11}(\infty)|$.  


On a straight domain wall in the $x$ direction, $m_{01}=m_{00} = 0$, while 
$m_{10}\neq0$.   The spectrum of midgap states depends on the
symmetry of the wall.  If the stripe is site-centered
(bond-centered), wall fermions have a gapless (gapped) spectrum.    
The absence of a gap can be ascertained by finding a zero mode at the momentum 
$p_x\equiv -i\partial_x=0$:  
\begin{equation}
\sigma_2 \frac{d\psi}{dy} = \frac{U}{2ta} 
[\tau^y_3 m_{11}(y) + i\tau^y_2 m_{10}(y)] \psi(y). 
\label{eq_0_modes}
\end{equation} 
To zeroth order in $m_{10}$, 
solutions of Eq.~(\ref{eq_0_modes}) are eigenstates of $\sigma_2$,
$\tau^y_3$ and, e.g., $\tau^x_3$ (each zero mode comes from a
single Fermi patch), giving a total of 8 linearly independent zero
modes.  As usual,\cite{Rajaraman} only half of these solutions [those
  with $\sigma_2\tau^y_3 m_{11}(+\infty)<0$] are localized on the
wall, so that there are 4 zero modes.  Remarkably, in addition to the
usual twofold spin degeneracy, there is another spin-like degree of
freedom, which will prove to be the transverse flavor.  The origin of
isospin (at weak coupling) is thus exposed: compared to a 1D chain,
there are twice as many ``Fermi points'' on a straight domain wall in
2D --- see Fig.~\ref{fig_dirac}, right.

The difference between one-particle spectra of 
site and bond-centered walls
arises in the first order in $m_{10}$.  Eq.~(\ref{eq_0_modes}) 
has four zero modes if $m_{10}$ is an odd function of $y$,
i.e., for a site-centered wall.\cite{Bloch}
On a bond-centered wall, $m_{10}(y)$ is even and fermions 
have a gap (Fig.~\ref{fig_dirac}):  
$$E(0) = \pm U \langle m_{1} \rangle
\equiv \pm U\int m_{10}(y) \psi^\dagger(y)\psi(y) dy.$$

The one-particle spectrum $E(p_x)$ on a straight site-centered 
stripe can be determined approximately by starting with $p_x=0$ 
[Eq.~(\ref{eq_0_modes})] 
and treating the first term in Eq.~(\ref{HF_continuum_2}) 
perturbatively.   In the limit $p_x\to 0$, states outside the main gap
can be neglected, which reduces the Hilbert space
to the four zero modes [Eq.~(\ref{eq_0_modes})].  
By using the degenerate perturbation theory, we find a Dirac spectrum 
(dashed lines in Fig.~\ref{fig_dirac}, left): 
\begin{equation}
E(p_x) \sim \pm v p_x, 
\hskip 3mm
v=2ta \langle\tau^y_1\rangle.
\label{H_eff_scw}
\end{equation}



{\em Domain-wall holons at $U\ll t$.} As illustrated in
Fig.~\ref{fig_site_to_bond}, magnetization on a {\em bond\/}-centered
wall with a wiggle can be obtained by superimposing $m({\bf r})$ of a
straight {\em site\/}-centered wall and that of a spin chain with a kink
in $x$-staggered magnetization.  Away from the wiggle, $m_{00}({\bf
  r})= m_{01}({\bf r}) = 0$.  To simplify the discussion, we will
neglect these components altogether.
Decompose $m({\bf r})$ into an $x$-independent
part and the rest:
\begin{eqnarray}
m({\bf r}) &=& m^{(0)}(y) + m^{(1)}({\bf r}),
\nonumber\\
m^{(0)}(-y) = -m^{(0)}(y),
&& 
m^{(1)}(\pm\infty,-y)=m^{(1)}(\pm\infty,y).
\nonumber
\end{eqnarray}
The Hamiltonian (\ref{HF_continuum_2}) can now be split in two parts:
\begin{eqnarray}
H^{(0)}_{\rm HF} = -2ita\, \sigma_1 \tau^y_3 \partial_y 
-U\sigma_3 [m^{(0)}_{11}(y) + m^{(0)}_{10}(y) \tau^y_1],
\label{H0}\\
H^{(1)}_{\rm HF} = -2ita\, \sigma_1 \tau^x_3 \partial_x
-U\sigma_3 [m^{(1)}_{11}({\bf r}) + m^{(1)}_{10}({\bf r}) \tau^y_1]. 
\label{H1}
\end{eqnarray}
As shown above, the ``transverse part'' (\ref{H0}) has 4 zero modes
for each $p_x$.  Within this Hilbert space, $H^{(1)}_{\rm HF}$ describes 
right and left-moving fermions with spin, which see a staggered magnetization
$$\langle m_{1}(x) \rangle = 
\int\!dy\ u^\dagger(y)
[m_{10}({\bf r}) + m_{11}({\bf r})\tau^y_1]u(y),$$
where $u(y)$ is a zero mode (\ref{eq_0_modes}) of Eq.~(\ref{H0}).  The
midgap fermion band acquires a gap of its own, 
$\tilde{\Delta} = U|\langle m_{1}(\infty)\rangle| < \Delta$,
with two zero modes (one for each spin) inside this smaller gap.
``Longitudinal'' wavefunctions of the two zero modes satisfy the equation 
\begin{equation}
\sigma_2 \frac{d\psi(x)}{dx} 
= \frac{U}{2ta\langle\tau^y_1\rangle}\,\tau^x_3 
\langle m_{1}(x) \rangle\, \psi(x).
\label{eq_0_modes_wall}
\end{equation}

The existence of two holon flavors can now be deduced from 
Eqns.~(\ref{eq_0_modes}) and (\ref{eq_0_modes_wall}).  The zero modes
have a finite norm only if
$$\sigma_2\tau^y_3 m^{(0)}_{11}(+\infty) < 0, 
\hskip 3mm
\sigma_2\tau^x_3 
\langle m_{1}(+\infty) \rangle / \langle\tau^y_1\rangle < 0.$$
It follows then that the product of eigenvalues 
\begin{equation}
\tau^x_3\tau^y_3 = 
{\rm sgn} [ m^{(0)}(y=+\infty) \, \langle m_1(x=+\infty)\rangle \,
\langle\tau^y_1\rangle ]
\label{signs}
\end{equation}
can be identified with the holon isospin $2\rho$.  This can be seen by
extrapolating Eq.~(\ref{signs}) to larger values of $U$, which reduces
the size of holons.  We have $\langle\tau^y_1\rangle =
\langle(-1)^y\rangle = (-1)^{y_0}$, where $y_0$ is the row number of
the chain in Fig.~\ref{fig_site_to_bond}.  According to
Eq.~(\ref{signs}), if $\tau^x_3\tau^y_3 = +1$, spins on the chain and
to the right (left) of the wiggle are an extension of the upper
(lower) AF domain, as for the $\rho=+1/2$ wiggle in
Fig.~\ref{fig_site_to_bond}.  Thus, $\rho = \tau^x_3\tau^y_3/2$.  This
identification is consistent with numerical HF solutions
(Fig.~\ref{fig_holon}),
where $\tau^x_3\tau^y_3 = {\rm sgn} k_x\, {\rm sgn} k_y$ can be inferred
from the orientation of a holon.

{\em Effective holon Hamiltonian}.
Large-$J_z$ cartoons (e.g., Figs.~\ref{fig_cartoon} and
\ref{fig_site_to_bond}) suggest that any configuration of a domain
wall is uniquely represented by an interface drawn through integer (if
a hole is present) or half-integer (bond, no hole) lattice points.  In
the absence of overhangs, the imaginary-time interface dynamics due to
spin exchanges and hole hops can be described by an SOS-type model.
(Locality in time assumes that coupling to the bulk spin waves can be
neglected.\cite{anisotropy}) In the dilute limit, mobile
single-particle excitations are spinons and holons, each equipped with
an isospin $1/2$.
When spinons are frozen out,\cite{anisotropy}
the only remaining excitations are holons, and 
the SOS model reduces to that of (iso)spin-$1/2$
fermions with the short-range Hamiltonian 
\begin{eqnarray}
  H&=&\sum_i\bigl[\bigl(
      -\tilde{t}\,\psi_{i\rho}^\dagger\psi_{i+1\,\rho}
      +  \tilde{J}_x\,s^+_i\,s^-_{i+1}+
      {\rm h.c.}\bigr)
      \nonumber\\ 
      & & \qquad
      +\tilde{U}\, n_{i\uparrow}\,n_{i\downarrow}+\tilde{V}\,n_i\,n_{i+1}
      + \tilde{J}_z\,s^z_i\,s^z_{i+1}
    \bigr].
  \label{eq:full-effective-hamiltonian}
\end{eqnarray}
Here $n_i\equiv n_{i\uparrow}+n_{i\downarrow}$ is the holon-number
operator, $n_{i\rho}\equiv \psi^\dagger_{i\rho}\psi_{i\rho}$, and
{\em iso\/}spin operators are $s^z_i\equiv
(n_{i\uparrow}-n_{i\downarrow})/2$ and $s^+_i\equiv
\psi^\dagger_{i\uparrow}\psi_{i\downarrow}$.
Taking $\tilde{U}\to+\infty$ excludes double occupancies, while isospin
exchange $\tilde{J}_x$ allows two holons on one side of the stripe to hop to
the other side.  By construction, we do not expect isospin-rotation
invariance; in general, $\tilde{J}_z\neq \tilde{J}_x$. 

Holons can be transported between different stripes without disturbing
the AF order only in pairs (with opposite isospins).
In contrast to the model of a quarter-filled chain\cite{Zaanen9804300}
with the same Hamiltonian (\ref{eq:full-effective-hamiltonian}), the
field $\psi^\dagger$ describes $Q=1$ 
particles, and the appropriate pairing operator $
\psi^\dagger_{i\uparrow}\,\psi^\dagger_{i+1\downarrow}$ has charge
$Q=2$.  This operator is a combination of the (iso)singlet (SP) and an
(iso)triplet (TP$_0$) pairing operators,
\begin{displaymath}
  {\cal O}_{\rm SP},\,{\cal O}_{{\rm TP}_0} = 
  (\psi^\dagger_{i\uparrow}\,\psi^\dagger_{i+1\downarrow}\mp
  \psi^\dagger_{i\downarrow}\,\psi^\dagger_{i+1\uparrow})/\sqrt2,
\end{displaymath}
and an instability in either of these channels could lead to global
(Cooper-pair) superconductivity, even though isospin is local to a
given stripe.

To analyze this scenario, we have performed a standard weak-coupling
analysis\cite{Emery-1DEG,Giamarchi-Schulz} at a generic density.
Because a stripe horizontal on average is invariant under isospin
reflection, isospin and charge degrees of freedom
separate.\cite{tilt-note} The scaling in each sector is determined by
the usual constants $K_s$, $K_c$; for repulsive interactions
$K_s>K_c^{-1}>1$.  In the absence of the isospin gap, both SP and
TP$_0$ correlation functions have the temperature
exponent\cite{Giamarchi-Schulz} $\mu_{\rm SP}=\mu_{{\rm TP}_0}
=K_c^{-1}+K_s-2>0$, i.e., there is no divergence in either
channel.  Unlike in a more conventional 1DEG,\cite{Emery-1DEG} where
the spin-rotation invariance requires $J_z=J_x$ and another mechanism
({\em e.g.,} the ``spin-gap proximity effect''\cite{spin-gap}) is
needed to develop the spin gap, here an isospin gap arises naturally
for $\tilde{J}_z>|\tilde{J}_x|$ (an easy-axis anisotropy).  At low
temperatures, such a system is in the Luther-Emery phase, the TP$_0$
component freezes out, while the isosinglet pairing exponent becomes
$\mu'_{\rm SP}=K_c^{-1}-2$; it diverges for $K_c>1/2$.  Isospins are
ordered at $T=0$ and the stripe becomes, on average, site-centered
(Fig.~\ref{fig_site_to_bond}, right).


As usual, the more divergent ($\mu_{\rm CDW}=K_c-2<\mu_{\rm SP}$)
CDW correlations compete with the superconducting order.  In our holon-only
model~(\ref{eq:full-effective-hamiltonian}), transverse stripe
fluctuations\cite{Kivelson-98} {\em never\/} suppress this
instability.  CDWs on neighboring stripes remain phase-coherent
not only in the isospin-ordered phase (where stripe
fluctuations are suppressed by the isospin gap), but also in the presence
of fully developed transverse fluctuations. 

Large-$J_z$ cartoons indicate that the length of a stripe seen by a
hole increases in the presence of spinons
(cf. Ref.~\CITE{Kivelson-98}).  This is why we believe that in the
full interface model (with both spinons and holons present, as in
Fig.~\ref{fig_cartoon}) the transverse fluctuations may reduce the
coherence of CDWs on neighboring stripes.  However, other properties
of this model require a separate study.


{\em Acknowledgments.}  The authors thank M.~M. Fogler, S.~A.
Kivelson, F. Wilczek and J. Zaanen for illuminating discussions.  This
work has been supported by DOE Grant DE-FG02-90ER40542.


\begin{references}
\bibitem{Neglect_dn} 
As charge order is driven by spin order in this case, density 
fluctuations do not play a crucial role.
\bibitem{anisotropy} Magnons and spinons are suppressed in the presence of
a spin gap (created, e.g., by an Ising anisotropy, $J_z > J_x$).
\end{references}
\end{document}